# Rapid Exploration of Assembly Chemical Space of Molecular Graphs

Ian Seet[1], Keith Y. Patarroyo[1], Gage Siebert[3], Sara I. Walker[2,3], Leroy Cronin[1]*

[1]School of Chemistry, University of Glasgow, Glasgow, G12 8QQ, UK.

[2]BEYOND Center for Fundamental Concepts in Science, Arizona State University, Tempe, AZ, USA

[3]School of Earth and Space Exploration, Arizona State University, Tempe, AZ, USA

*Corresponding author email: Lee.Cronin@glasgow.ac.uk

**Abstract**

Quantifying how hard it is to build a molecular graph matters for biosignature detection, chemical complexity, and cheminformatics. We present an exact, scalable algorithm to compute the molecular assembly index (MA) which prioritizes the largest duplicate subgraphs, represents fragmentation with an "assembly state" array of edge-lists, reuses states via hashing/DAGs, and prunes the search using a dynamic-programming branch-and-bound guided by a conditional-addition-chain lower bound. For organic molecules in the greater than 500 Da range our approach is up to six orders of magnitude faster than prior methods and yields exact MAs where previous algorithms would have timed out. We compute MAs to convergence for ~300k COCONUT natural products with <50 bonds, profiling time and memory scaling. Finally, we exploit the speed of our algorithm to calculate joint assembly spaces and introduce the Joint Assembly Overlap (JAO), a Jaccard-like metric that emphasizes global scaffold reuse and show that the JAO yields substantially different rankings from Tanimoto similarity with ECFP fingerprints and MCS (e.g. in steroids 270–380 Da and short peptides), accounting for substructural similarity beyond local environments. Together, these advances turn the molecular assembly index into a practical tool for large-scale exploration of chemical space.

## Introduction

Cheminformatics hinges on quantifying structure, similarity, and complexity to explore the vastness of chemical space. Standard tools to quantify similarity e.g.— the fingerprint-based Tanimoto metrics (e.g., ECFP-4/6)[1] and maximum common substructure (MCS)[2]—are fast and effective, but the former emphasises local neighbourhoods at the cost of global features, while the latter only considers the largest common substructure, presenting problems when the targets have many disjoint common substructures. Assembly theory offers a different perspective: the molecular Assembly Index (MA) measures the minimal informational constraints needed to build a molecular graph—the fewest joining steps required when previously assembled fragments can be reused (illustrated for benzoic acid in Fig. 1 on p. 2, MA = 6). A molecular graph is an abstract representation of the structure of a chemical compound. The *Assembly Index* of a molecule attempts to capture the minimal informational constraints needed to construct such abstract representation. As such it is defined as the fewest number of steps required to make its molecular graph by recursively using previously made structures. For example, we might consider the following molecule (benzoic acid), with its bonds as building blocks, Figure 1a.

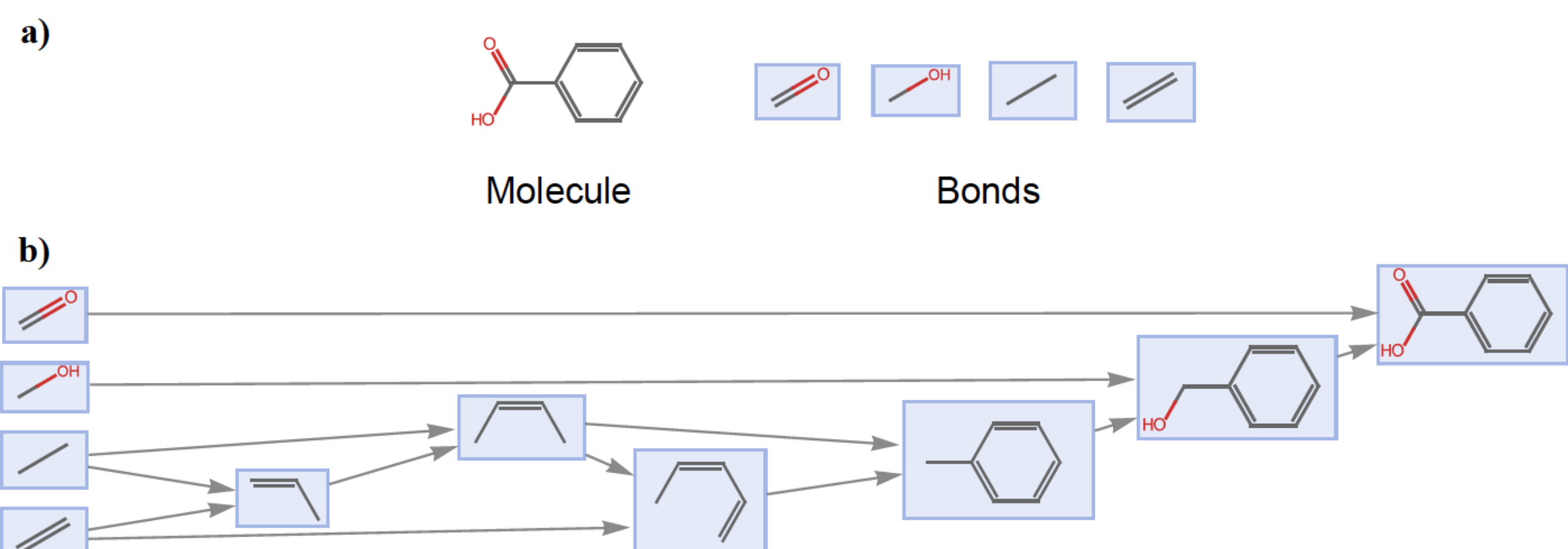

**Fig. 1:** a) Molecular graph of benzoic acid, with the set bonds being used to represent it. b) Progressive minimal construction of the molecule benzoic acid from a set of bonds, at each step a pair of structures are joined or “glued” like Lego pieces in order to construct the desired molecule.

We can obtain the assembly index by counting the steps in a minimal way to construct the molecule using shared vertex assignments to join bonds and intermediate structures, Figure 1b. The assembly index of a molecular graph was first proposed by Marshall et al[3] in the context of finding biosignatures in the search for life in other planets. It has received recent attention for the exploration of chemical space[4], the measurement of chemical complexity[5], and the quantification of evolution and open-endedness[6, 7].

It would initially appear that a key problem linked to finding the assembly index of a molecular graph is that of molecular subgraph enumeration. Enumerating connected subgraphs has been used to define the complexity of molecular graphs[8] and for graph substructure mining with the aim of designing molecular graphs with specific properties[9]. Typically, this enumeration is performed in a depth-first manner[10], and some canonical ordering is induced depending on the specific application. While our algorithm enumerates all possible duplicate subgraphs once at the start of the algorithm, we subsequently store the relationships between the duplicated subgraphs as a directed acyclic graph which we re-use. As we do not repeat the enumeration process, this subgraph enumeration is not generally the slowest step of the algorithm.

The problem of finding minimal addition chains[11] is equivalent to finding the assembly index of a chain of bonds with one bond and atom type. Addition chains and their properties are extensively studied problems of which one of its generalizations have been proven to be NP-Complete[12]. This last problem viewed in the context of assembly indexes and assembly spaces is equivalent to find the joint assembly space[7] of a set of chain of bonds with one bond and atom type. Moreover, as highlighted in Marshall et al[13], both the minimal addition chain and vectorial addition[14] chain are special cases of the formalism of assembly spaces and are employed to compute lower bounds of the assembly index of more complex spaces.

The process of calculating the assembly index of a molecular graph can generate a compressed representation which can be sent through a communication channel, albeit at considerable computational cost. One can alternatively represent a molecule using a molecular specification format, like SMILES[15] and compress the resulting set of strings with a text based compression technique[16, 17]. These techniques may generate more optimal compression for strings, but they lack the structural properties of the assembly construction process. Furthermore, there is little evidence that they perform well for compressing molecular graphs[18]. On the other hand the problem of context-free grammar based text compression, while also being computationally costly, shares a similarity with assembly pathways in the hierarchical nature of the representation of the compressed objects[19, 20].

Other than molecular graphs, one can compute assembly indices of other data structures like strings, pixelated images and voxelized 3D objects[13]. These data-structures have compression techniques that take advantage of the sparse nature of the objects[21-23], this resembles the way in which the assembly index construction recursively uses redundant data. If one ignores the specific nature of the data-structures, one could resort to universal sequence data compression techniques[16]. While they may provide a considerable compression ratio, it differs in its structural nature to that of the assembly index construction process and once again do not appear to be effective for molecular graph compression[18]. Our initial algorithms[5] have used nauty[24, 25] in the canonical labelling of enumerating all possible duplicatable subgraphs. While nauty is a fast graph isomorphism[26] library, it is a general graph isomorphism algorithm and furthermore does not explicitly handle edge colourings, forcing additional vertices to be added to simulate edge colourings. Although there exist algorithms that can solve for molecular graph isomorphism in polynomial time[27], these algorithms are difficult to implement.In this work we mostly consider the case of molecular graphs arising from organic molecules, where there are relatively few duplicatable cyclic subgraphs and the maximum degree is low. We thus combine a strategy of tree isomorphism[28] for acyclic subgraphs and a general graph isomorphism using the VF2 library[29] for the rest. As graph isomorphism is generally not the slow step in the assembly algorithm, we do not believe it is practically necessary to implement the polynomial time isomorphism algorithms.

Several measures of complexity for molecular graphs have been proposed which attempt to capture structural properties of the graph[8, 30-32]. In particular, various measures have been proposed concerning the number of subgraphs in the molecular graph, these are indices which sum vertex degrees in subgraphs from the molecular graphs[30], count all the number of subgraphs[8] or spanning trees[31]. Such measures differ from the assembly index by being tied to the specific application for which they were developed. Also of importance are the measures of algorithmic information theory[33] such as Kolmogorov[34]. These measures attempt to quantify complexity in a universal senseby finding the shortest computer program that can produce the molecular graph. Although this is a very powerful measure, it is incomputable, unlike the assembly index. Differences between assembly index and computational complexity measures were expanded in Kempes et al[35].

Our earlier research has tackled the computation of the assembly index of molecular graphs[3-5] and strings[6, 7]. Since the duplicate subgraph enumeration is very computationally intensive for large molecules, earlier approaches approximated the process by splitting the molecular graph in large substructures with little overlap[3]. Other approximations rely on random sampling of duplicate subgraphs for molecules[4] and a binary tree decomposition for strings[6]. The efficient computation of an exact assembly index of molecules [5, 13] and strings[7] has been explored recently with a depth-first subgraph enumeration with a logarithmic branch and bound. In this work we build on this work and introduce a dynamic programming with a sophisticated branch and bound to compute assembly indices efficiently of large molecules.

**Definitions**

A Let $G_M = (V, E)$ be a *molecular graph*, where $V$ is the set of vertex associated with atoms of the molecule excluding hydrogen atoms. The set $E$ is the set of edges associated with bonds, e.g. covalent, organometallic, ... that can be found between the atoms of the molecule. The association of vertex and edges with atoms and bonds is given by the labelling functions $l_V: V \to \Sigma_V$ and $l_E: E \to \Sigma_E$, where $\Sigma_V$ and $\Sigma_E$ represent the type of atoms and bonds present in the molecule respectively. Given a molecule or in general a set of molecules $\{G_M^i\}_{i=1}^n$, an *assembly* construction process, constrained by all the atom and bond types $\Sigma_V^T, \Sigma_E^T$ of the set, is a construction procedure generated by a set of objects, called virtual objects or fragments and a set of joining operations that build fragments from other fragments.

**Definition 3.1:**

The set of *virtual objects* or *fragments* $\Omega$ represent all molecular graphs with fixed atom and bond types $\Sigma_V^T, \Sigma_E^T$ determined by a fixed set of molecular graphs $\{G_M^i\}_{i=1}^n$.

With this definition in mind, we now consider how to build fragments in this space.

**Definition 3.2:**

Given two fragments $x, y \in \Omega$ we define the *joining operation* such that $x \odot y = z$ is the molecular graph resulting of the union of $x$ and $y$ plus an identification of a non-trivial set of vertices of $x$ and $y$ that are made identical.

This formalizes the idea of "gluing" two molecular graphs by a specific set of common atoms. Note that there are multiple ways of joining $x$ and $y$, the notation $x \odot y = z$ in this case means that there exists a joining operation which combines $x$ and $y$ into $z$.

An important subset of objects that are useful for building fragments is the set of *building blocks*,

$$B_M = \{u \in \Omega : |E_u| = 1\},$$

therefore, the building blocks are the set of molecular graphs in $\Omega$ with one bond.

Whenever we are performing an assembly construction process, we accumulate a set of fragments. We formalize this with the following definition,

**Definition 3.3:**

An *assembly pool* $P$ is any set of fragments $P \subset \Omega$ such that for all $z \in P \backslash B_M$, $\exists x, y \in P$ such that $x \odot y = z$.

We can now consider what is the minimum number of steps to build a fragment $x \in \Omega$ or a set of fragments $X \subset \Omega$ that are contained in a specific assembly pool $P$.

**Definition 3.4:**

The *assembly index* of a set fragments $X \subset \Omega$ is the minimum size assembly pool that contains $X$ excluding the building blocks,

$$a_i = \min_{P(X)} |P(X) \backslash B_M|,$$

where $P(X)$is an assembly pool that contains $X$. Given a specific assembly pool $P(X)$, we can generate an assembly space, where we define it as,

**Definition 3.5:**

An *unlabelled assembly space* $\Gamma$ generated by an assembly pool $P(X)$ is a multi-directed acyclic graph(multi-DAG) where we have $x \in P \Leftrightarrow x \in V_\Gamma$ and for $x, y, z \in P, x \odot y = z \Leftrightarrow [x, z], [y, z] \in E_\Gamma$.

Note that since the graph is a multi-DAG, we can have $(x, z)$ repeated twice meaning that $x$ was used twice to build $z$.

If we add an edge-labelling map $\phi: E_\Gamma \to V_\Gamma$ such that $\phi([x, z]) = y$ and $\phi([y, z]) = z$, then we have an equivalent definition of an *assembly space* to the general quiver formulation in[13]. In this paper we refer to the unlabelled assembly space as assembly space for simplicity. We shall also refer to the assembly space of a set of fragments $X$ as a *joint assembly space*. Finally, we will refer to an *assembly path* or *assembly pathway* as a specific topological ordering of an assembly space's vertices.

Although we aim to find the shortest pathway required to construct a graph from two-bond fragments, the graph assembly algorithm does not directly calculate the assembly index by searching through the space of possible pathways. Rather, it attempts to find duplicatable subgraphs in the molecule-graph and iteratively removes subgraphs, Figure 2.

At each step of the process, we find a duplicatable subgraph within the molecule graph. We then remove this duplicatable subgraph from the structure by deleting all edges (but not nodes) from the original graph and disconnect the replicated structure associated to the duplicatable subgraph from the remnant structure. We continue this process until no possible duplicatable subgraphs remain.

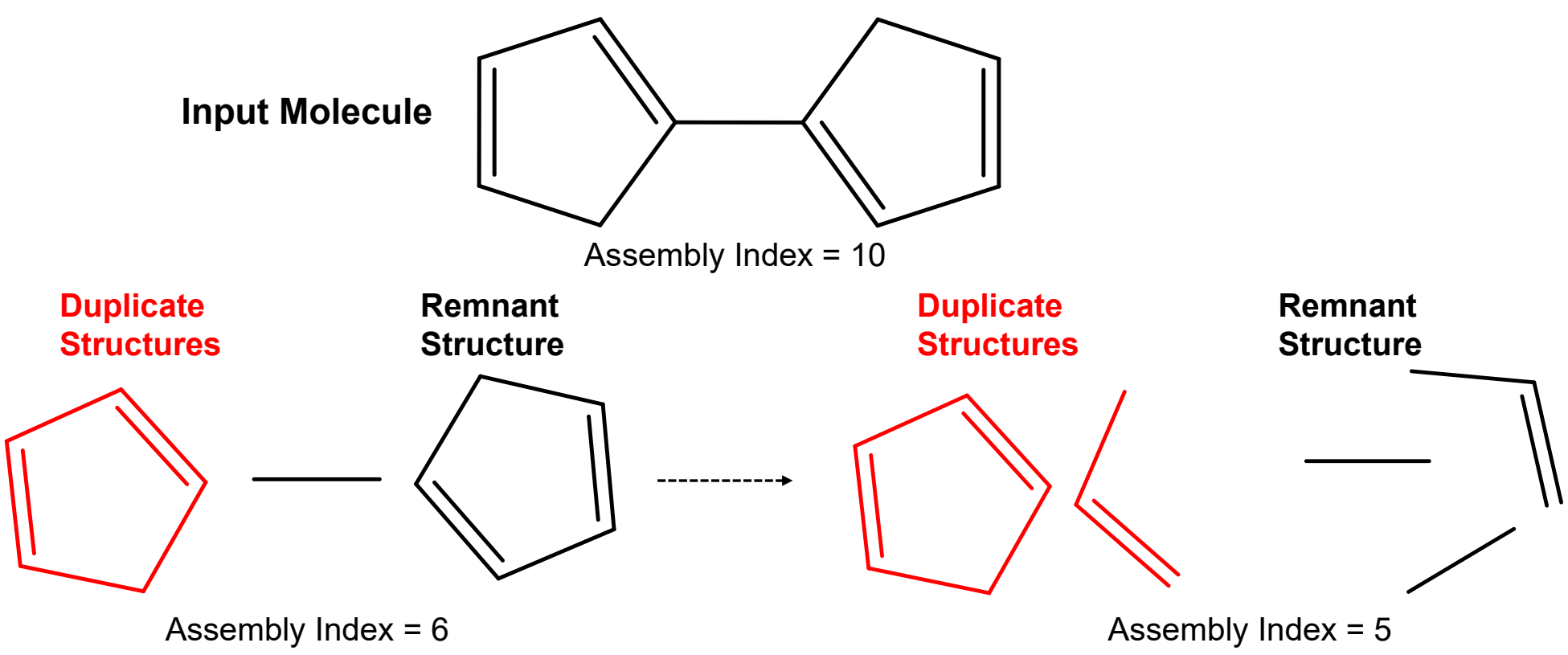

**Fig. 2:** Fragmentation of a molecule-graph via iteratively searching for and removing duplicatable subgraphs.

The maximum possible assembly index for a graph is $N - 1$, where $N$ is the number of edges in the graph. Such a graph has no duplicatable subgraphs with at least two bonds. In order to calculate the assembly index using this duplicate-finding method, for each duplicatable subgraph deleted, we subtract from the $N - 1$ upper bound the value $k - 1$ where $k$ is the number of edges in the subgraph. This is because each duplicate of size $k$ represents a saving of $k - 1$ bonds over a one-bond fragment during the forward assembly process. The sum of all $k - 1$ values for all possible duplicates $S$ can be subtracted from $N - 1$ to obtain the assembly index for a particular molecule.

## Data Structures

As all possible fragments generated by the aforementioned process must necessarily be fragments of the original molecule-graph, we may efficiently represent such fragments as a *boolean edgelist*. A boolean edgelist is an array of boolean variables of size $N$ where $N$ is the number of edges in the original molecule-graph. Each boolean variable within the array corresponds to the presence or absence of a particular edge for a given fragment. This boolean edgelist may be conveniently implemented as a bitset variable in C++.

Each step in the assembly process may be stored as an *assembly state*. An assembly state consists of an array of edgelists, where the first element is the edgelist of the last fragment found in the previous state. The fact that the first element in the edgelist is the last element taken can be used to ensure that pathways containing permutations of identical duplicatable fragments are not investigated more than once. An assembly state also stores the value of $S$, the duplicate sum of the pathway that resulted in this specific collection of fragments. This duplicate sum can be used to find a lower bound on the assembly index that would be obtained for a given pathway.

**Algorithm**

The graph assembly algorithm seeks to find the shortest assembly pathway and does so by finding the pathway with the largest value of $S$. Broadly, it does so by first finding all possible duplicatable subgraphs in the initial graph. It then iteratively generates assembly states by removing the duplicatable subgraphs from the parent graph until no possible duplicatable subgraphs remain. For most molecule graphs, the duplicatable subgraph enumeration is not generally the slowest step; rather, it is the iterative fragmentation to find the pathway with the maximum value of $S$ which is the most time-consuming part of the algorithm.

We implement three major heuristics to reduce the time complexity of the iterative fragmentation. Firstly, we iterate through duplicatable subgraphs in inverse order of size such that the largest duplicates are processed first. Secondly, we utilise dynamic programming by hashing and storing all assembly states to prevent the algorithm from processing states it has encountered before unless those states have a higher value of $S$. Thirdly, we implement a branch and bound heuristic where we exploit the fact that the duplicates are searched in reverse order of size to establish a tight lower bound on the maximum obtainable value of $S$ for a given assembly pathway. In addition to these major heuristics, we implement several other techniques to prune the search tree which we shall explain in greater detail in the following sections.

The first step in the graph assembly algorithm is to delete all unique bonds as they cannot possibly be part of any duplicatable subgraph. This algorithm is trivially of time complexity $O(V + E)$ where $V$ is the number of vertices and $E$ the number of edges of the molecular graph and only needs to be performed once at the start of the algorithm. Subsequently, all potential duplicated matching substructures are enumerated and hashed. The pseudocode for this portion of the algorithm is as follows:
Subsequently, all potential duplicated matching substructures are enumerated and hashed (Algorithm 1). The function `Matching_Enumerator()` is used to generate a new set of duplicatable subgraphs from a previous set by adding all possible adjacent edges to the graph via the function `Duplicate_Generator()`. We use a persistent variable GlobalHashMap, a hash map with a graph key and an integer value to index all distinct subgraphs encountered. If two graphs are isomorphic, they will have the same key-value pair. We store all isomorphic sets of subgraphs in the local hash map LocalHashMap, which has an integer key and a value corresponding to a list of isomorphic subgraphs. By collect all isomorphic subgraphs into a single list, we can find all potential matching subgraphs by comparing their boolean edgelists to detect overlaps via the function `Matching_Validity()`, with non-overlapping boolean edgelists constituting valid matches.

While there exist methods to perform molecular graph isomorphism in guaranteed polynomial time[27], for the purposes of our algorithm, our inputs are largely organic molecules which do not generally have a large number of cyclic duplicatable subgraphs. In particular, many duplicatable subbgraphs will be acyclic. Tree isomorphism can be implemented in $O(N)$ time where $N$ is the number of nodes of the trees to be compared. Constructing an adjacency list from an edgelist and checking if a graph is cyclic can trivially be performed in $O(N)$ time, thus, if the duplicatable subgraphs are acyclic, isomorphism can be performed in linear time.

In the case where the duplicatable subgraphs are not acyclic, we use the C++ graph canonization algorithm implemented in the VF2 library[29] to check if cycle-containing subgraphs are isomorphic. Although the worst-case performance for the general graph isomorphism problem is not provably of polynomial time complexity, we find that VF2 is fast in practice when executed on most molecular graphs. For the duplicate enumeration performed on every assembly state apart from the original molecule, we keep track of the sorted index of the most recently removed duplicate (i.e. the first element in the assembly state's list of edgelists as mentioned in the section on data structures). To prevent the algorithm from evaluating multiple permutations of the same assembly state, we only evaluate duplicates which have a sorted index smaller than or equal to the most recently removed duplicate. This restriction on maximum duplicate size also allows for a tighter lower bound to be calculated for the assembly index of a given assembly state, as we elaborate in the next section.

During the enumeration, we keep track of every bond which is part of every potential duplicatable subgraph with a bitset variable, where we initialise a bitset variable to 0 and set the bit corresponding to the bond's index in the edgelist to 1 if the bond exists as part of any duplicatable subgraph. If a bond is not part of any duplicatable subgraph, we remove it from consideration in a process directly analogous to the preprocessing step.

On the first pass of the algorithm, we organise the duplicates into a directed acyclic graph (DAG). The nodes of this graph correspond each unique duplicatable subgraph and each directed edge points from a given duplicatable subgraph to every subgraph with one additional edge that was discovered during the enumeration process. Construction of this DAG can be accomplished in $O(V_d + E_d)$ time where $V_d$ are the number of unique duplicatable subgraphs and $E_d$ the number of potential edges of the DAG. Note that although a given duplicatable subgraph can be constructed from multiple parent subgraphs, we only maintain an edge from the first such parent encountered, minimising the size of the DAG. Furthermore, we do not have to repeat this step on subsequent passes of the algorithm as the DAG is preserved.

```
Function Duplicate_Generator (TargetGraphs, LocalHashMap):
          for g ∈ TargetGraphs do
             for e ∈ AdjacentEdges do
                NewGraph := g.append(e)
                if GlobalHashMap.contains(NewGraph) is false then
                    increment Key by 1
                    GlobalHashMap.insert(Key, NewGraph)
                 end
                n := GlobalHashMap.keyOf(NewGraph)
                LocalHashMap.valueOf(n).append(NewGraph)
          end
   end
Function Matching_Validity (g_1, g_2):
      if g_2.edgelist ∨ g_1.edgelist ! = 0 then
            RETURN false
      else
            RETURN true
      end
Function Matching_Enumerator (MatchingList, PriorDuplicates):
      Duplicate_Generator(PriorDuplicates, LocalHashMap)
      for v ∈ LocalHashMap.Values do
            for g_1 ∈ v do
                  for g_2 ! = g_1 ∈ v do
                        x := Matching_Validity (g_1, g_2)
                        if x == true then
                              MatchingList.append (g_1, g_2)
                              active := true
                        end
                  end
            end
      end
```

**Algorithm 1**: Subgraph matching enumeration.

**Branch and Bound Heuristic**

Before an assembly state is fragmented, it is reasonable to produce a crude lower bound of the minimum achievable assembly index for that state and delete this state from the stack should this lower bound be greater than the lowest assembly index found thus far. A trivially provable lower bound is $\lceil \log(N) \rceil$ where $N$ is the total number of bonds in the assembly state, but we can achieve a much tighter lower bound by exploiting the fact that all children of a particular assembly state will have a maximum duplicate size not greater than the parent due to the duplicate evaluation heuristic described in the previous section.

This bound on the maximum size of the largest duplicatable subgraph which may be taken for all children of a particular assembly state can in turn be used to bound the value of $S$ (and thus, the minimum obtainable assembly index) by the following expression:

$$S = \max(b(x): x = 2,3,..,m-1,m)$$

$$b(x) = -\lceil \log_2 x \rceil + \sum_i L - \lceil L/x \rceil$$

where $m$ is the maximum size of the largest duplicatable subgraph and $L$ is the size of each fragment in the assembly state up to the $i$th fragment.

In order to rigorously prove this expression, we define a new problem: the conditional addition chain problem. This is a variant of the addition chain problem but where at a specific integer $m$ must be used and no number larger than this integer may be used except in combination with a number smaller than or equal to this integer. This is because there is a limit on the maximum size on the largest duplicatable subgraph which corresponds to the limit on the size of $m$. Thus, the length of the shortest path solution to this problem is therefore also a lower bound to the analogous graph problem.

We shall now prove that this expression yields a maximal value of $S$ for a conditional addition chain, and thus also represents a lower bound on $S$.

**Lemma 4.1** For a conditional addition chain with specific integer $m$ and size $l$, the shortest chain length cannot be smaller than is $\lceil l/m \rceil + \lceil \log_2 m \rceil - 1$.

**Proof:** If $l$ is divisible by $m$, it is trivial to see that that the number of addition steps not including the steps required to construct $m$ is $\frac{l}{m} - 1$. Since $m$ must be used at least once regardless, the term $\lceil \log_2 m \rceil$ which describes the minimum addition chain length required to construct $m$ must be included.

In the case where $l$ is not divisible by $m$, there is no benefit to including a sum where the smaller number is smaller than $m$ more than once, corresponding to the remainder of $\frac{l}{m}$. This is because $\frac{l}{m-x}$ where $x > 0$ must necessarily be greater than $\frac{l}{m}$. Thus, the number of addition steps required not counting the steps required to construct $m$ is $\lceil l/m \rceil - 1$. □

**Lemma 4.2** For a conditional addition chain of a set of integers of sizes $\{l_1, l_2 \ldots l_n\}$ with specific integer $m$, the shortest chain length cannot be smaller than $\sum_i \lceil l_i/m \rceil + \lceil \log_2 m \rceil - 1$.

**Proof:** We can apply lemma 4.1 individually to each integer $l_i$ in turn since each integer's addition chain is independent of the others, with the exception that the $\lceil \log_2 m \rceil - 1$ term need only be used once.

**Lemma 4.3** The optimal pathway for a conditional addition chain of a set of integers up to a maximal specific integer $m$ cannot be shorter than the conditional addition chains for specific integers $2,3,..,m - 1, m$ for any $l \geq 2$.

**Proof:** From lemma 4.1. □

From lemmas 4.1 and 4.3, we may now derive equation (2). The upper bound on the value of $S$ is equal to the lower bound on the assembly index $A$ from Lemma 4.1 subtracted from the maximal assembly index of a fragment $L - 1$ for a given $m$, see (4). From (4), we may derive (2) by using lemmas 4.2 and 4.3.

$$L - 1 - A = S$$

$$S = L - \lceil L/m \rceil - \lceil \log_2 m \rceil$$

It is possible to replace the $\lceil \log_2 m \rceil$ term with pre-calculated assembly indices for uniform linear strings of length $m$, but the effects on time efficiency are not significant, with the improved lower bound possibly being outweighed by poorer cache performance.

The simple conditional addition chain heuristic may be extended further by exploiting the fact that it is rare for $L$ to be equal to the maximum number of bonds $L_m$ that may be taken for fragment size $m$. Thus, one can calculate separate conditional addition chains for $L - L_m$ with fragment size $m - 1$, and $L_m$ with fragment size $m$. The sum of the two separate conditional addition chains may then be used in place of $S$.

## Fragmentation

As we have now established an ordering of duplicatable subgraphs, we must now remove the duplicatable fragments from the list of boolean edgelists. This may seemingly be accomplished by a simple binary bitwise XOR operation between the boolean edgelist of the duplicatable fragment and the boolean edgelist of the parent. However, a complication may arise if this removal causes the parent to further fragment into several smaller graphs. Directly hashing the resulting edgelist is not desirable as there are far more combinatorial possibilities if the edgelist is hashed directly than if it is fragmented first. Furthermore, as the branch-and-bound heuristic we use benefits greatly from having each distinct fragment enumerated, we require an efficient algorithm to separate the remnant boolean edgelist corresponding to the former parent graph into a list of edgelists, each corresponding to the graph of each connected fragment.

```
Function Disjoint_Set_construction (DisjointSet, EdgeList) :
        for p ∈ EdgeList do
                if  DisjointSet.find (p.first)  is false  then
                        DisjointSet.insert (p.first)
                end
                if  DisjointSet.find (p.second)  is false  then
                        DisjointSet.insert (p.second)
                end
                DisjointSet.union (p.first, p.second)
        end
Function Disjoint_Set_Splitting(DisjointSet,  AssemblyState) :
        for i ∈ DisjointSet do
                DisjointSet.compress(i)
        end
        for i ∈ DisjointSet do
                HashMap.valueOf(DisjointSet.find(i)).append(i)
        end
        for j ∈ HashMap.values do
                AssemblyState.append(j)
        end
```

**Algorithm 2**: Disjoint-set construction and splitting.

We may accomplish this task by using a modified disjoint-set data structure to reconstruct all connected subgraphs for a given edgelist (Algorithm 2). To do so, we first run the disjoint-set construction function on the target boolean edgelist and then use the disjoint-set splitting function on that disjoint set. The result of this operation is a set of boolean edgelists corresponding to the remaining fragments. The time complexity of this algorithm is $O\big(E\ \alpha(E)\big)$ [36] where $E$ is the number of edges in the list of edgelists and $\alpha$ is the inverse Ackermann function; this function is practically linear in $E$. We then apply the branch-and-bound heuristic a second time on the fragments produced by the disjoint-set splitting.

**Assembly State Hashing**

The boolean edgelist obtained from the disjoint-set reconstruction is then sorted in $O\big(N\log(N)\big)$ time; the comparator function used for the sorting can be arbitrary as long as it is consistent. We subsequently append the original fragment to the set to create a list of edgelists where the original fragment is the first element, which corresponds to the assembly state previously mentioned in the section on data structures. Since each boolean edgelist has a unique hash value, we may use the edgelist hash function to convert the list of edgelists into a vector of integers, which may in turn be trivially hashed in average case $O(N)$ time using any string hashing algorithm. By hashing each assembly state we prevent identical states from being evaluated more than once unless there is an improvement in the sum of duplicate bonds found.

**Recursion and Pathway Generation**

With the branch and bound heuristic and the assembly hash table, we may eliminate the majority of assembly states produced by the fragmentation step from consideration. We then recursively evaluate the remaining states in descending order of $m$. This arrangement results in the states with the largest values of $m$ being evaluated first, which intuitively results in a good upper bound early in the execution and improves the ability of the branch and bound algorithm to eliminate states which cannot reach this bound.

Recovering the assembly pathway can be accomplished by taking advantage of the fact that all unique assembly states are stored in a hash table. We retain a pointer between each assembly state and its immediate parent. Should an assembly state have its value of $S$ updated, we replace the original pointer with a pointer to the parent of the state which triggered the update. Thus, we can reconstruct the pathway by taking the pointer of the assembly state with the maximum value of $S$ and iterate through the parents of each pointer until we arrive at the original assembly state. From this state we obtain the duplicate and remnant structures mentioned in the section on data structures. The procedure to reconstruct a specific minimal pathway is described in the Appendix A.

**Benchmarking**

In order to assess the performance of the proposed algorithm, we consider several test cases to illustrate the advantages of this paper's algorithm when compared to prior state of the art. We start by considering a test of 12 molecular graphs relevant in different areas of chemistry, see Table 1.

| **Molecule** | $a_i$ | **Bonds** | **Depth-First** | | **This Work** | |
|---|---|---|---|---|---|---|
| | | | Time(s) | Memory(MB) | Time(s) | Memory(MB) |
| SR1001 | 22 | 31 | 1444 | 40.0 | 0.006 | 6.82 |
| Quinoline Yellow | 11 | 24 | 1027 | 184.3 | 0.021 | 7.35 |
| Dienogest | 11 | 26 | 3089 | 156.8 | 0.050 | 9.27 |
| Pirenperone | 19 | 32 | >3600 | - | 0.051 | 7.38 |
| Ketoconazole | 22 | 40 | >3600 | - | 0.225 | 8.39 |
| Cefpirome | 25 | 39 | >3600 | - | 0.113 | 8.44 |
| Cefiderocol | 30 | 54 | >3600 | - | 4.71 | 25.0 |
| Cefpimizole | 27 | 50 | >3600 | - | 3.00 | 24.7 |
| Tetranactin | 9 | 60 | >3600 | - | 102.6 | 2678 |
| Phosphatidylcholine | 22 | 55 | >3600 | - | 4.51 | 21.5 |
| Erythromycin | 20 | 53 | >3600 | - | 15.05 | 57.7 |
| Iodotaxol | 20 | 50 | >3600 | - | 20.71 | 11.1 |

**Table 1:** Memory and time comparison of state-of-the-art methods for calculating the assembly index of molecular graphs.

From this set, only three have an assembly index that can be calculated exactly with previous methods[5]. Our method computes the assembly index up to six orders of magnitude faster. Furthermore, for the remaining nine molecules, the assembly algorithm described in this paper is able to compute all the assembly indices where prior state of the art can only deliver an approximation of the assembly index when allotted an hour of computing time.

From these molecules we can observe the progressive calculation of the assembly index, as exemplified with the molecule Iodotaxol, Figure 3a. This molecule is the anti-cancer drug Taxol, but with the phenyl groups replaced by an iodine atom. Figure 3a. We find that the algorithm finds the correct assembly index, $a_i = 20$ in less than 1 second, with the remaining time being spent on verifying the correctness of the solution. We also show a reconstructed minimal pathway, Figure 4 from the duplicate and remnant fragments, Figure 2. In order to reconstruct this pathway from the algorithm output we used the procedure described in the Appendix A.

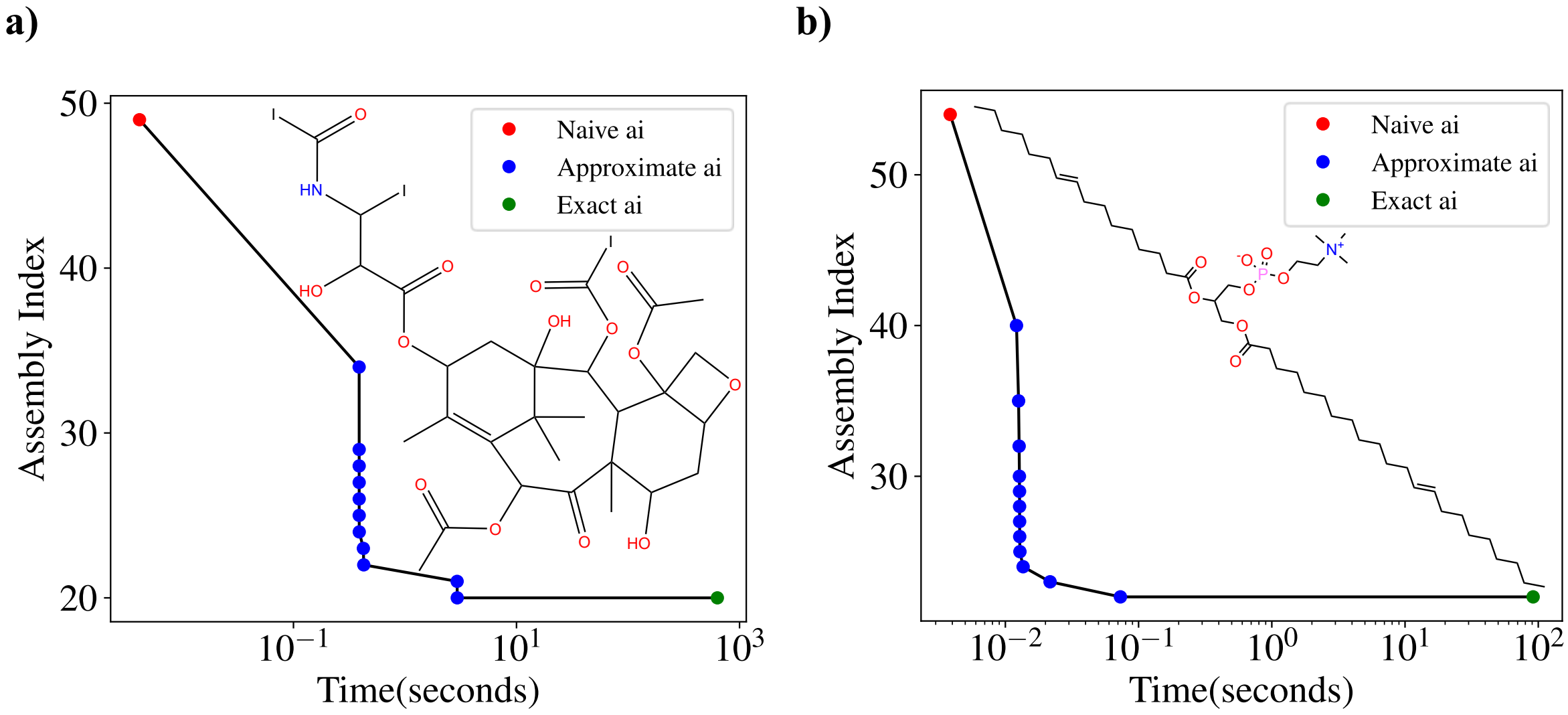

**Fig. 3:** a) Progressive approximation of the assembly index from the molecular graph of Iodotaxol given computational time until convergence to the exact value. b) Progressive approximation of the assembly index for the molecular graph of Phosphatidylcholine given computational time until convergence to the exact value.

For the remaining test cases, we consider molecules with a total number of bonds from 20 to 60. Within this range our algorithm is capable of finding the assembly index in less than a minute, over a wide variety of substructural motifs. The algorithm also functions well for largely linear molecules such as Phosphatidylcholine (Fig. 3b). This molecule i consists of a long linear backbone with a small sidechain that contains just over 50 bonds in total, with three bond types. As before, the exact assembly index is calculated in a matter of milliseconds, and it takes less than five seconds to recurse over the rest of the search tree to confirm the solution.

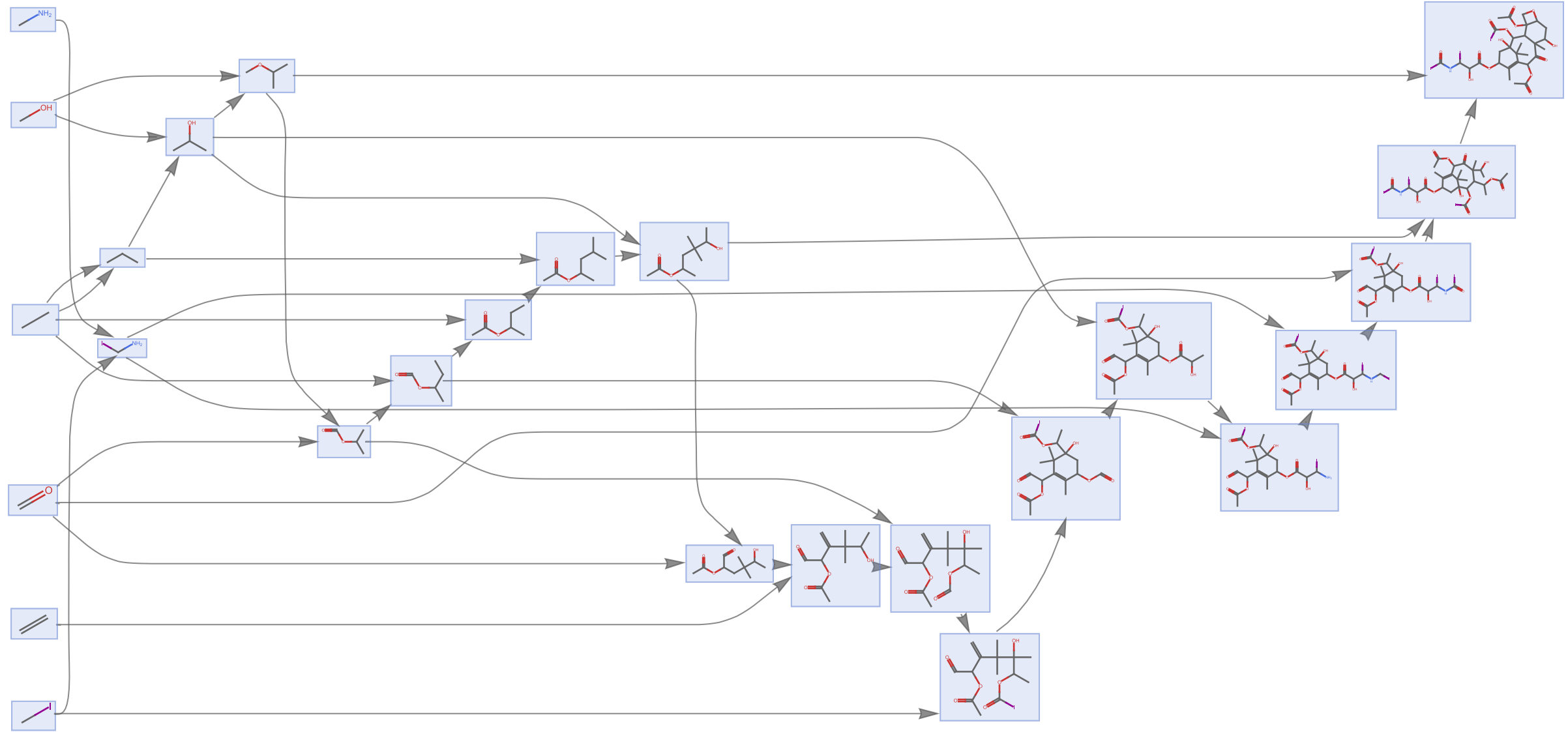

**Fig. 4:** Pathway reconstruction of the molecular graph of Iodotaxol from the duplicated and remnant structures produced by the algorithm.

Linear chains are also of computational interest, because if we restrict the number of atom types and bond types we can enumerate all possible molecules up to a certain length. If we consider a linear chain with only one type of atom, and only two types of bonds, we can enumerate all possible molecules, which can be mapped to binary numbers, Figure 5a. We computed the assembly index of all assembly indices up to length 20 with a previous method[5] and up to length 25 with our method. We can clearly see that our algorithm is both faster by a constant factor and demonstrates better asymptotic scaling than the previous best assembly algorithm.

Our algorithm is also naturally capable of computing joint assembly spaces. In contrast with the standard molecular assembly index, the joint assembly index is calculated formultiple disjoint molecular graphs. The joint assembly index is the minimum number of joining operations simultaneously construct the set of disjoint molecules. To test the capabilities of our algorithm, we consider the set of all standard 20 amino acids. With the current capabilities of our algorithm, we are able to compute the exact joint assembly index of about 13 amino acids; for a larger number of amino acids, we can stop the algorithm early and obtain an approximation. We compute the joint assembly index of all 20 amino acids for one hour and obtain the approximation $a_i = 37$, see Figure 5b.

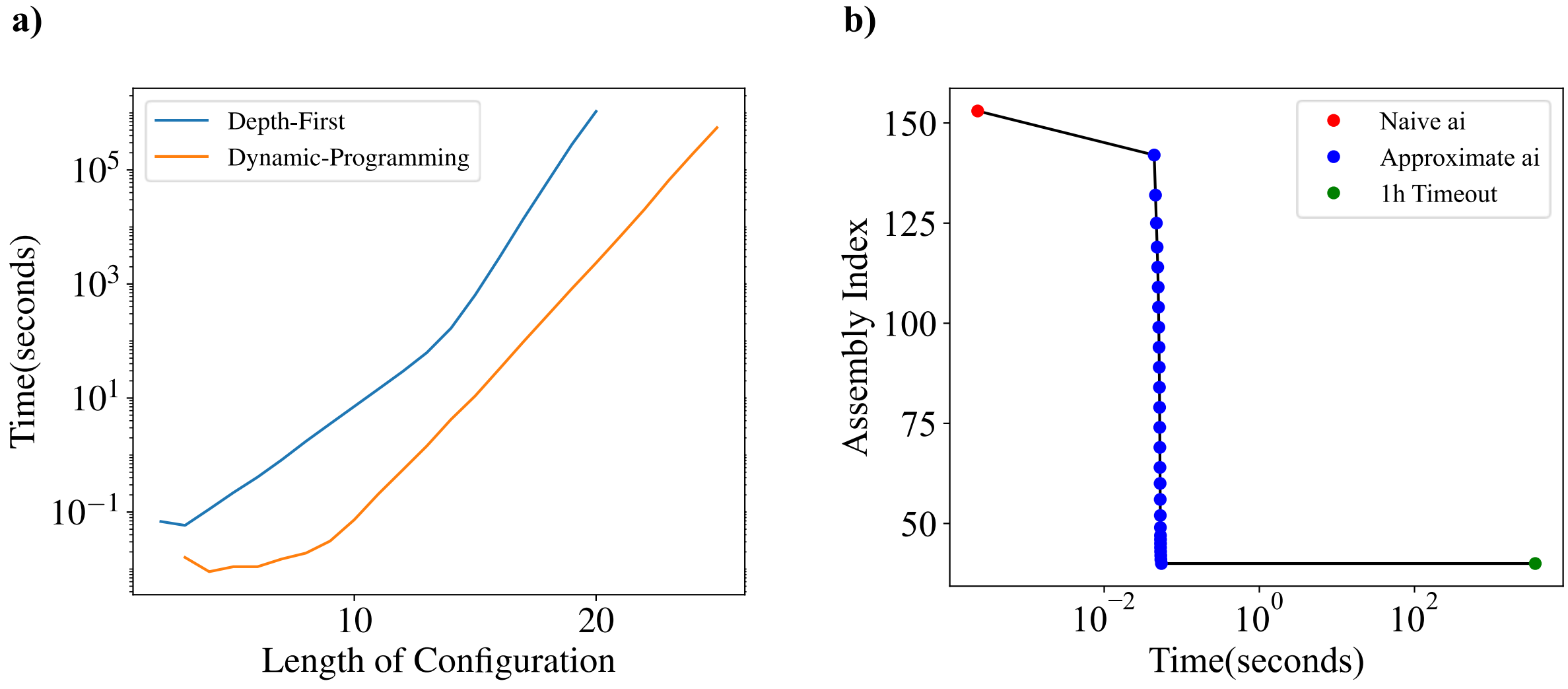

**Fig. 5:** a) The scaling curves for the computation of the total amount of time needed to calculate all possible configurations assembly indexes of linear two-bond molecule chains of length $n$ and one atom type. Shown are the dynamic programming approach described in this paper and a naive depth first approach with a simple $\log N$ branch and bound heuristic described in[5]. b) Progressive approximation of the assembly index the joint assembly index of the combined molecular graph for standard 20 amino acids given computational time until a timeout of one hour.

To probe how much reusable structure exists among biogenic building blocks, we computed a joint assembly space for the 20 standard amino acids by treating their molecular graphs as a disconnected input and then reconstructing a minimal shared pathway from the algorithm's duplicate/remnant output. The resulting space exposes the expected common motifs, e.g., the α-amino/α-carboxylate backbone together with small alkyl and aromatic fragments that are assembled once and then reused across multiple amino acids. Quantifying reuse by comparing the joint assembly index of the set with the sum of the individual MAs we obtain ≈70% compression relative to a naïve construction with no reuse of fragments between individual molecules. In other words, roughly two-thirds of the joining operations required to build each amino acid independently are eliminated when shared substructures are propagated through the joint space ($MA_{joint}$ ≈ 37 after 1 h). This shared pathway provides a compact, interpretable dictionary of fragments and joins that both summarizes biochemical regularities and serves as a lossless, assembly-aware representation for large molecular sets. Because the representation is an assembly space (a DAG of fragments and joins), it is naturally easier to query and supports downstream tasks such as similarity (Joint Assembly Overlap, JAO), clustering, and incremental updates, while preserving chemically meaningful structure—advantages that generic string/graph compressors do not provide.

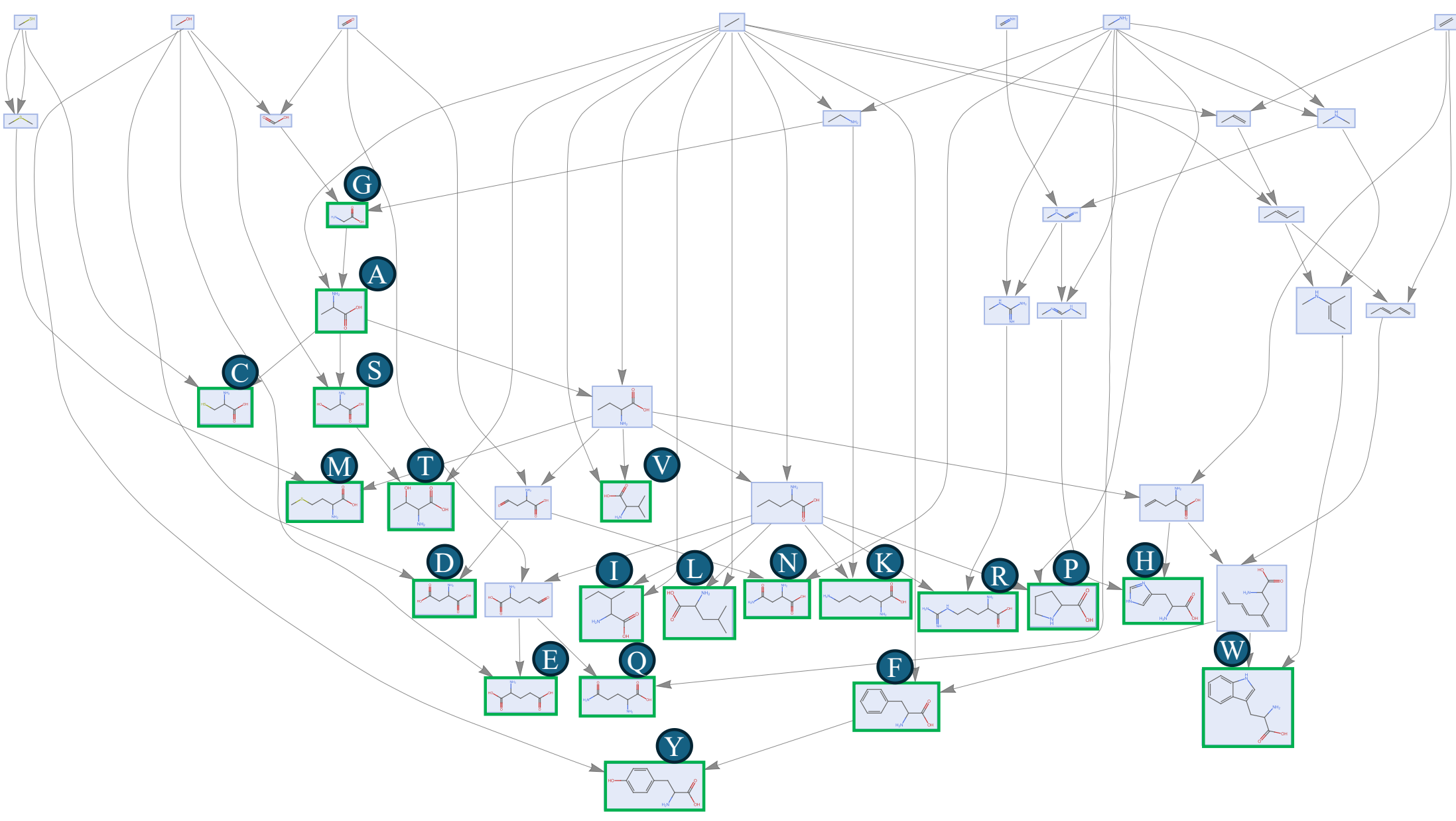

**Fig. 6:** Pathway reconstruction of the combined molecular graph of all standard 20 amino acids from the duplicated and remnant structures produced by the algorithm.

Subsequently, we took all molecules from the COCONUT database[37] with less than 50 bonds, equalling a total of about 300 000 molecules and computed their assembly index until convergence. We kept track of the total amount of time needed for convergence and the maximum memory usage needed for computation. The results are shown in Figure 7. In both graphs, we have grouped the molecules by number of bonds. We clearly see that the amount of time and memory needed to compute a molecule with specified assembly index grows exponentially.

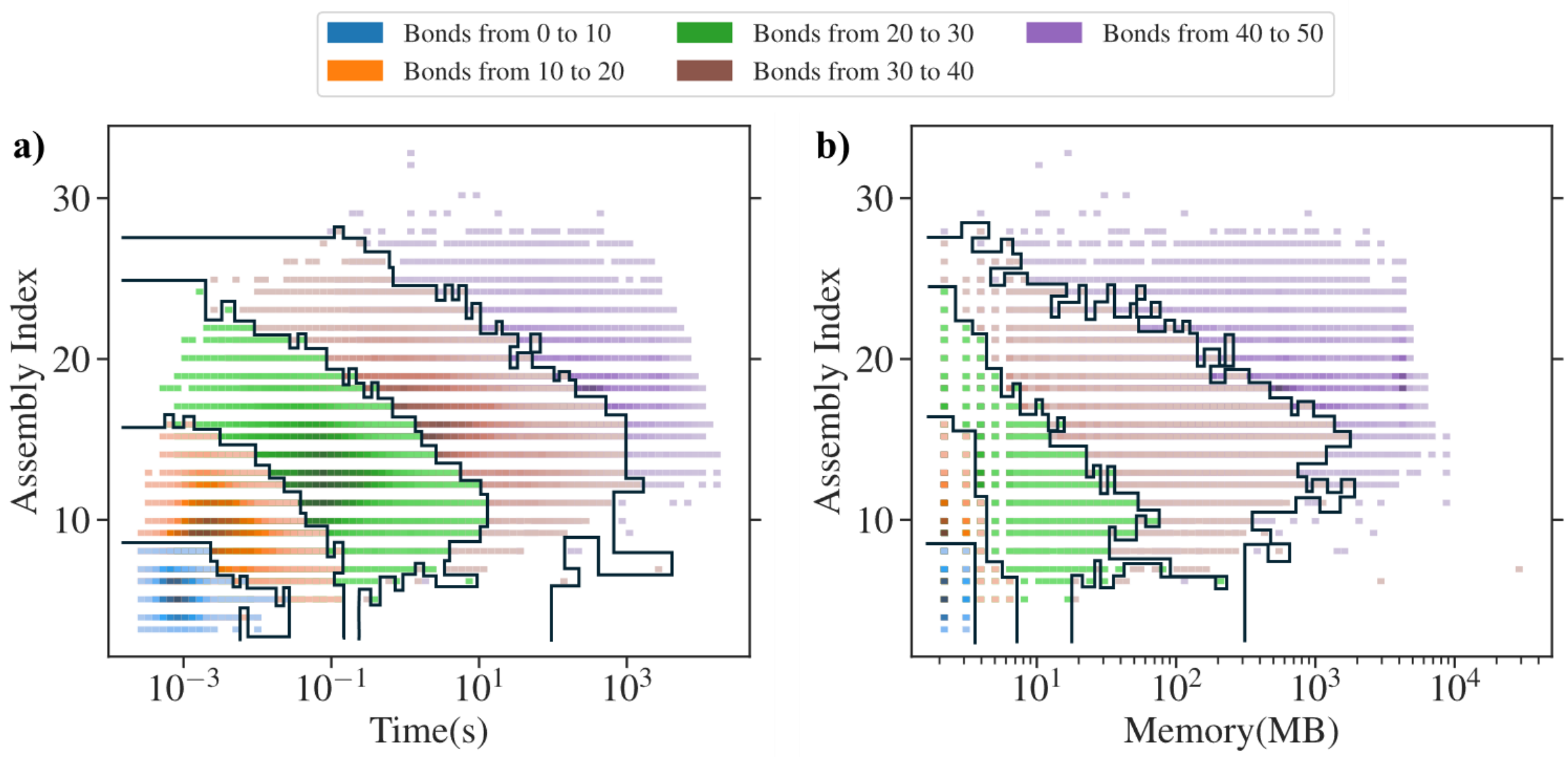

**Fig. 7:** a) Distribution of assembly index versus total time of computation of a dataset of around 300 000 of molecules coming from the COCONUT database. The molecules are grouped by total number of bonds. b) Distribution of assembly index versus maximum memory usage needed for computation of a dataset of around 300 000 of molecules coming from the COCONUT database. The molecules are grouped by total number of bonds.

The increased speed of our assembly algorithm has lent itself to a variety of applications that would not have been possible with prior algorithms, including exploring biochemical space[38], training a machine learning model to estimate MA from mass spectra[39], and quantifying the complexity of crystalline materials[40]. Beyond these applications, we shall show that the increase in speed also allows the MA to be used as a similarity metric that has substantially different properties from other commonly used molecular similarity metrics.

**Joint Assembly Index as a Similarity Metric**

Unlike most other molecular complexity metrics such as the well-known Bertz[41], Bottcher[42], Whitlock[43] or Spacial[44] scores, the Assembly Index calculated for disjoint sets of molecules (henceforth referred to as joint assembly spaces) is not merely the sum of the assembly indices of each individual molecule within the space. Instead, the joint Assembly Index accounts for common substructural information between molecules in the joint assembly space and is thus smaller for molecules with significant substructural overlap.

By treating the Assembly Index of a joint assembly space as the Assembly Index of the union of the individual assembly spaces, a Jaccard Index-like metric analogous to the Tanimoto similarity metric can be calculated, henceforth referred to as the Joint Assembly Overlap (JAO):

$$JAO_{A,B} = \frac{MA_A + MA_B - MA_{A,B}}{MA_{A,B}}$$

Where $MA_A$ and $MA_B$ are the assembly indices of compounds A and B respectively and $MA_{A,B}$ is the joint assembly index of both A and B. The JAO has been described before[45], but only in the context of MAs estimated via mass spectrometry rather than exact MAs calculated from molecule graphs.

Although superficially similar in form to the Tanimoto similarity metric, we can show that the JAO is generally more sensitive to global symmetries over local symmetries when compared to the ECFP-4 and ECFP-6 fingerprints most commonly used with the Tanimoto similarity metric. It is also more capable of detecting similarities between disjoint common subgraphs than the MCS similarity metric. In the example illustrated in Fig. 8, the Tanimoto similarity of the two molecules under the ECFP-4 and ECFP-6 fingerprints is 0.231 and the MCS similarity is 0.333, but the equivalent JAO is 0.667. This is because the assembly algorithm can use both FCCBr and ICCCl fragments independently in the construction of the assembly pathway.

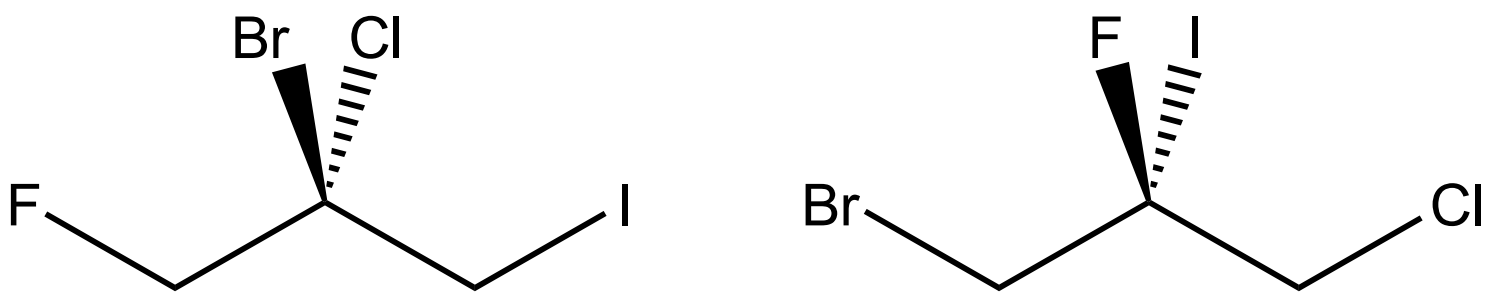

**Fig. 8:** A pair of compounds where the Tanimoto and MCS similarity metrics diverge substantially from the JAO.

The example illustrated in Fig. 8 highlights the important point that the JAO can account for substructural similarity beyond what the ECFP fingerprints and MCS scores can detect. There exist classes of molecules with similar skeletons but different local environments, such as steroids (see Figure 9). The greater efficiency of our assembly algorithm allows us to find the JAOs (or at least good estimates thereof) of these relatively large organic molecules (270-380 Daltons) where previous assembly algorithms would have been unsuccessful.

Here, the JAO tends to yield a higher similarity than the Tanimoto similarity metrics, while still being smaller on average than the MCS. Note that although the JAO and the MCS both rely on finding common graph substructures, one is not a monotonic function of the other as can be observed by the MCS of Cholesterol (1) and Estradiol (3) being substantially smaller than the JAO of the same compounds when it is greater for most of the other pairs of compounds.

| Combination | ECFP-4 | ECFP-6 | MCS | JAO |
|---|---|---|---|---|
| 1 + 2 | 0.207 | 0.142 | 0.429 | 0.375 |
| 1 + 3 | 0.257 | 0.178 | 0.256 | 0.333 |
| 1 + 4 | 0.3 | 0.219 | 0.462 | 0.462 |
| 1 + 5 | 0.354 | 0.25 | 0.447 | 0.462 |
| 2 + 3 | 0.203 | 0.15 | 0.486 | 0.375 |
| 2 + 4 | 0.438 | 0.33 | 0.897 | 0.75 |
| 2 + 5 | 0.459 | 0.341 | 0.828 | 0.615 |
| 3 + 4 | 0.286 | 0.188 | 0.485 | 0.267 |
| 3 + 5 | 0.418 | 0.321 | 0.567 | 0.462 |
| 4 + 5 | 0.723 | 0.638 | 0.852 | 0.636 |

**Fig: 9.** Tanimoto similarity metrics with ECFP-4 and ECFP-6 fingerprints, MCS similarity and JAO values for combinations of steroids Cholesterol (1), Cortisol (2), Estradiol (3), Progesterone (4) and Testosterone (5).

While we have illustrated examples of the JAO being lower than ECFP Tanimoto similarity for pairs of molecules, there are examples for which the reverse is also true. In particular, small peptides with individual amino acids transposed will in general have higher ECFP Tanimoto similarities than JAO scores due to the largely similar local environment, see Table 2.

| Combination | ECFP-4 | ECFP-6 | MCS | JAO |
|---|---|---|---|---|
| ACD + ADC | 0.879 | 0.652 | 0.667 | 0.538 |
| EFG + EGF | 0.702 | 0.565 | 0.562 | 0.647 |
| HIK + HKI | 0.862 | 0.704 | 0.806 | 0.765 |
| LMN + LNM | 0.905 | 0.7 | 0.724 | 0.625 |
| PQR + PRQ | 1 | 0.886 | 0.75 | 0.647 |
| STV + SVT | 0.879 | 0.727 | 0.826 | 0.5 |
| WAY + WYA | 0.86 | 0.744 | 0.641 | 0.737 |

**Table 2**: Tanimoto similarity metrics with ECFP-4 and ECFP-6 fingerprints, MCS similarity and JAO values for tripeptides containing all 20 standard amino acids. Each tripeptide is represented by the one-letter abbreviations of each amino acid.

## Conclusions

In this paper we have introduced a novel molecular-graph assembly index algorithm, specifically designed for organic molecules with few cycles and low maximum vertex degree. We have described the algorithm in detail, highlighting the different stages in its implementation. We have performed an extensive experimental evaluation on a set of relevant examples in the chemistry literature and a database of natural products, comparing the execution time of our algorithm and other recent assembly index algorithms. The results of the experimentation confirm that our algorithm has a reasonable memory usage even for on molecular graphs of substantial size, and its execution time exhibits several orders of magnitude of improvement with respect to previous methods such that for organic molecular graphs, our algorithm is the fastest existing assembly index algorithm. The increase in speed has allowed the assembly algorithm to be used for a variety of tasks (exploring biochemical and crystalline material space and generating a sufficiently large dataset to train a machine learning model) not otherwise practical with prior algorithms. Furthermore, it also allows us to use our algorithm as a similarity metric, the Joint Assembly Overlap (JAO) for relatively large organic molecules. We show that as a similarity metric the JAO differs substantially from the two of the most commonly used similarity metrics in cheminformatics, the Tanimoto similarity metric with ECFP-4 and ECFP-6 fingerprints and the MCS similarity metric due to its ability to account for disjoint global substructural similarities.

**Supplementary Data and Code Availability**

All the code required for assembly calculations and generating the figures are available at https://github.com/croningp/assemblycpp-v5. All the data required to produce the figures is available as a supplementary data file and a supplementary document explains how the figures were made.

**Acknowledgements**

The authors would like to thank Stuart Marshall for comments and suggestions early in the manuscript, and Amit Kahana for the idea of benchmarking joint assembly spaces by adding one molecule at a time. We acknowledge financial support from the John Templeton Foundation (grant nos. 61184 and 62231), the Engineering and Physical Sciences Research Council (EPSRC) (grant nos. EP/L023652/1, EP/R01308X/1, EP/S019472/1 and EP/P00153X/1), the Breakthrough Prize Foundation and NASA (Agnostic Biosignatures award no. 80NSSC18K1140).

**Author Contributions**

L.C. conceived of assembly theory and developed it with S.I.W. I.S. developed the algorithm and the similarity metric analysis. K. Y. P. did the benchmarking and the visualizations and G.S. developed the mathematical formalism. K.Y. P, I.S., and L.C. wrote the manuscript with input from all the authors.